**The societal impact of ion beam therapy**

Thomas R. Bortfeld[1*], Matthew Fernandez de Viana[2], Susu Yan[1]

[1]Department of Radiation Oncology - Division of Radiation Biophysics
Massachusetts General Hospital and Harvard Medical School
100 Blossom St
Boston, MA 02114, USA

[2]The University of Western Australia
35 Stirling Highway
PERTH WA 6009, Australia

[*]Corresponding author: tbortfeld@mgh.harvard.edu

The COVID-19 pandemic has changed the world and has diverted much of our public health capacity. This situation will likely be temporary and other global challenges, in particular cancer, will come back into focus. However, we expect that there will be a lingering effect of the crisis that, among many other consequences for the society, will cause financial scrutiny especially when it comes to big investments and costly technology. Denounced as breathtakingly expensive, ion beam therapy (IBT) will likely come under extra scrutiny. Voices critical of IBT will get louder. People will raise anecdotal concerns over particle therapy centers that have become white elephants, i.e., expensive to maintain and impossible to get rid of. More than ever it will be necessary to demonstrate a quantifiable societal benefit of IBT.

A widely supported health economic metric to quantify disease burden is the disability adjusted life year (DALY) [1]. The WHO defines DALY as a summary measure that combines time lost through premature death defined as years of life lost (YLL) and time lived in states of less than optimal health defined as the years lived with disability (YLD) [2]. The age-standardized cancer death rate has declined globally by 15% (20% in Germany) between 1990 and today [3]. As the survival rate of cancer is increasing because of earlier and better diagnosis, and better treatment options, we need to consider YLD more than YLL. Since the disability associated with radiotherapy is largely due to damaging healthy tissues, IBT becomes now even more relevant to improving DALYs [4,5,6,7].

In principle, if IBT would be available to every cancer patient, this treatment could greatly reduce cancer burden in terms of DALY. However, this ideal scenario has not eventuated because IBT comprises complex and costly systems that require further technical development. A lot of research and clinical development has already been performed over the past 60 years. In the near future, the following strategies could accelerate and demonstrate a substantial benefit from IBT.

First, because the delivery of IBT is resource intensive, data-driven patient selection maximizes the impact. A canonical example of IBT is for pediatric tumors where long-term treatment side effects are minimized because of the physical advantage of ion beams [7]. Another applicable criterion is when tumor hypoxia becomes significant, resistance to other forms of radiation therapy can be overcome with the high linear energy transfer (LET) of heavy ions, which is not dependent on free radical formation [8]. Through biological modeling to predict the synergistic effects between immunotherapy and particle therapy, certain patients should be identified to benefit from these combined treatments [9] [10]. In addition, dosimetric models predicting patient specific benefits from IBT should be used in the clinic [11].

The second strategy for preventing DALYs is to increase the clinical utility of IBT through innovation in delivery techniques. Modern advancements in Medical Physics, and adjacent fields, have given us the ability to reduce treatment margins by minimizing range uncertainties [10] and utilizing real-time adaptive planning [14,15]. The notable lack of improvement in lung cancer outcomes with IBT [16] is likely attributable to a lack of precise tumor tracking and delivery techniques. The development of motion management in IBT could improve the dose conformity, approaching what fundamental physics allows. Another development in this category is FLASH therapy with ultra-high dose rates of 60 Gy per second or higher [17]. Recent studies in FLASH using proton beams have shown promise in pre-clinical studies [18]. If the underlying mechanism of the FLASH effect is indeed localized oxygen depletion [17], the FLASH effect in high LET IBT will be diminished. Many questions are still left to be answered, especially whether there is a unique FLASH effect in proton therapy that can be observed in humans, and if a clinical trial warrants the benefits.

Finally, the third strategy is to democratize IBT – by which we mean to make IBT affordable and accessible [19]. In contrast to our first strategy above, here the DALY averted per patient may seem relatively small but could add up to reduce global DALYs by millions. An example from conventional radiation therapy is intensity modulated radiation therapy (IMRT). The total number of patients treated with IMRT worldwide is on the order of 30 million[1]. By the conservative assumption that every IMRT patient had 1/12 DALY (i.e., one moth) averted due to IMRT's dosimetric advantage, the total DALY reduction is still impressive at approximately 2.5 million. It is safe to assume that if IMRT were as expensive as IBT, the success of IMRT would never have happened. However, it is worth noting that reduction of future costs to treat radiation side-effects can make IBT more justifiable.

The democratization of IBT therefore requires significant cost reduction, which can be accomplished through a combination of faster delivery (not necessarily at FLASH speeds), better system integration, more efficient workflows, and hypo-fractionated treatment regimen for shorter treatment duration. Substantial contributing factors to the overall cost are the initial capital cost of the equipment and the infrastructure cost, which scale with the system size. Several companies have spearheaded the development of "compact" systems with a full or half gantry. At the Massachusetts General Hospital, we have taken this idea to the next level by retrofitting a new proton therapy system into two bunkers designed for conventional linear accelerators (LINACs), such that no separate building was needed. The accelerator (a synchrotron) is housed in one of the former LINAC rooms. The 190-degree gantry fitted into the other room, after excavating a deep pit. One goal for the future is to miniaturize gantries or to remove them altogether [20] such that a (proton) system will eventually fit in a single unmodified LINAC room (Figure 1). This compact gantry-less system can be achieved by the development of three main components:  a horizontal fixed beam-line with a robotic chair/immobilization system for patient positioning, a fast imaging system to achieve on-line adaptation and position monitoring, and adaptive delivery of pencil-beam scanning. Such a system would significantly reduce the cost of proton systems, while achieving the same, or even better treatment accuracy. Clinical experience and research efforts of Medical Physicists in collaboration with industrial partners will facilitate the development of such a system.

---

[1] Estimate based on 14.1 million patients per year newly diagnosed with cancer, 1/3 of them get radiation, and 1/2 of those get IMRT. IMRT started in 2000 with linear ramp up between 2000 and 2015.

If more developments focus on the societal benefit of IBT through the three strategies above: patient selection based on clinical and biological models, innovative delivery technology, and democratization of the system for substantial cost reduction, we expect that IBT will manifest a highly significant societal impact by reducing the global burden of cancer.

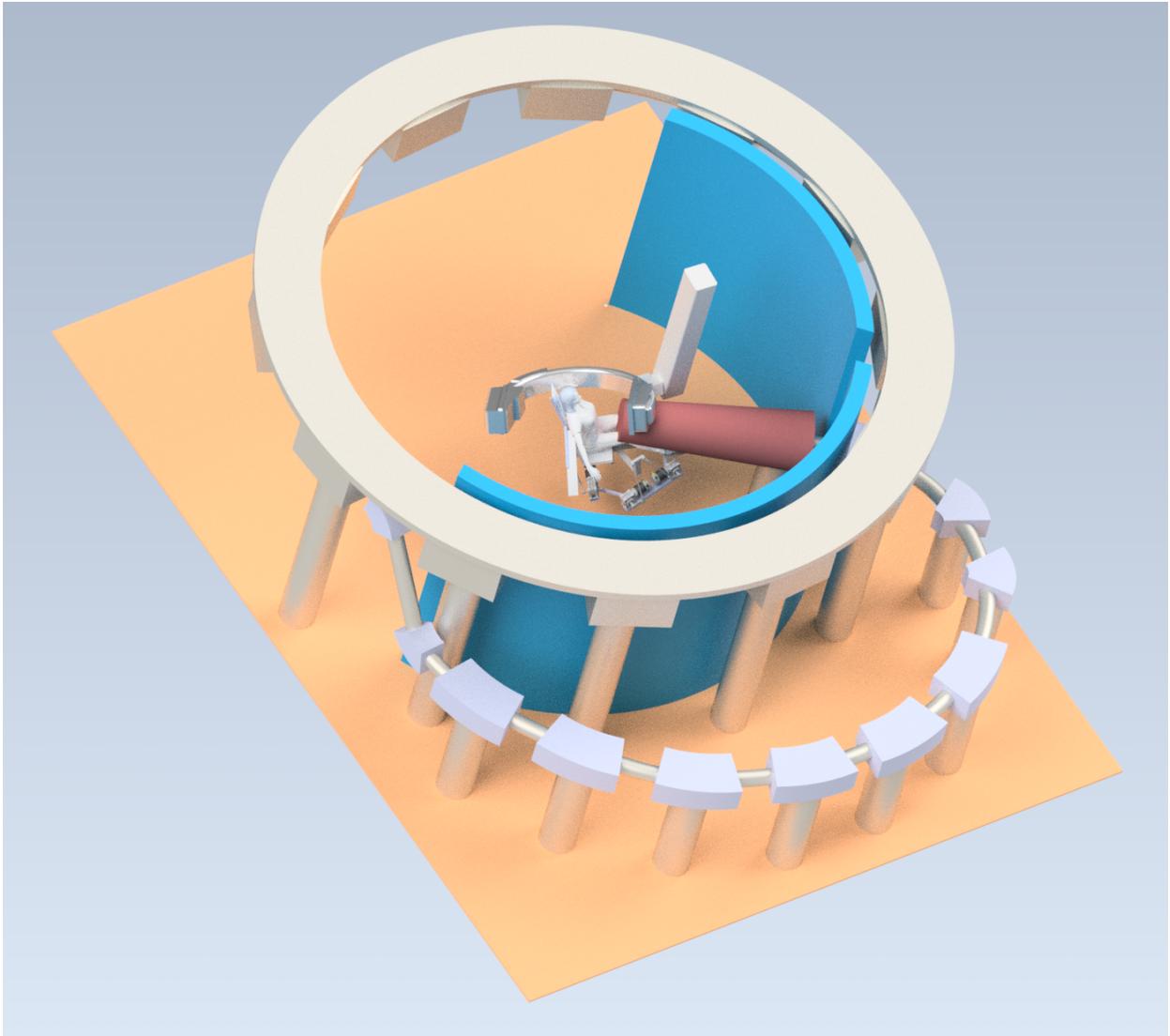

Figure 1: Schematic of a compact gantry-less single room proton system that can be retrofitted in a LINAC sized room (3D model by Fernando Hueso Gonzalez).